\documentclass{PoS}

\title{Phase quenching in finite-density QCD: models, holography, and lattice}

\ShortTitle{Phase quenching in finite-density QCD: models, holography, and lattice}

\author{\speaker{Masanori Hanada}\\
        KEK Theory Center, High Energy Accelerator Research Organization (KEK)\\
        E-mail: \email{hanada@post.kek.jp}}

\author{Yoshinori Matsuo\\
        KEK Theory Center, High Energy Accelerator Research Organization (KEK)\\
        E-mail: \email{ymatsuo@post.kek.jp}}

\author{Naoki Yamamoto\\
        Yukawa Institute for Theoretical Physics, Kyoto University\\
        Institute for Nuclear Theory, University of Washington\\
        Maryland Center for Fundamental Physics, Department of Physics, University of Maryland\\
        E-mail: \email{nyama@umd.edu}}

\abstract{
Finite-density QCD is difficult to study numerically because of the sign problem. 
We prove that, in a certain region of the phase diagram, the phase quenched approximation is exact to $O(N_f/N_c)$. 
It is true for any physical observables.  
We also consider the implications for the lattice simulations and find a quantitative evidence for the validity of the phase quenching from existing lattice QCD results at $N_c=3$.  
Our results show that the phase-quench approximation is rather good already at $N_c=3$, and the $1/N_c$  correction can be incorporated by the phase reweighting method 
without suffering from the overlap problem. We also show the same equivalence in effective models and holographic models. 
}

\FullConference{The 30th International Symposium on Lattice Field Theory\\
                 June 24 -- 29,  2012\\
                 Cairns, Australia}

\begin{document}

\section{Introduction}
QCD at a finite baryon chemical potential and/or finite temperature is an important subject of study, 
which is crucial for understanding the early universe, the relativistic heavy ion collisions, 
and the dense matter inside the neutron stars. 
Although the lattice QCD simulations should play an important role for studying the strongly coupled parameter region 
of this theory,  the notorious {\it sign problem} prevents us from a direct application of the simulation. 
In principle, a lattice simulation can be performed by using the phase-quenched ensemble. The effect of  the phase can be taken into account 
by the phase reweighting method. However whether it is practical or not is not clear a priori; when the number of the flavors $N_f$ is two, 
the phase quenched theory is the QCD with the isospin chemical potential\footnote{
As we will see in Sec.~\ref{QCDBvsPhaseQuench}, although this statement is correct for the partition function, 
there is a difference when one considers the physical observables.   
}, whose phase diagram is different from the original theory 
(for example the pion condensation takes place), and hence a severe overlap problem can appear. 

Recently it turned out that the phase quenching and the phase reweighting are actually practically useful techniques. 
The first to emphasized this fact, albeit empirically, are probably Kogut and Sinclair \cite{Kogut:2007mz}. 
They pointed out that various model calculations give the same answer for certain observables in the full and phase-quenched theories, 
as long as the pion condensation does not take place in the latter. They also pointed out that known lattice data give very similar results. 
Independently, Cohen \cite{Cohen:2004mw} and Toublan \cite{Toublan:2005rq} pointed out the similarity in the large-$N_c$ limit. 
Recently these facts have been understood theoretically unified manner \cite{Cherman:2010jj,Hanada:2011ju,Hanada:2012es}. 
Actually there is an exact {\it equivalence} between the full and phase-quenched QCD at large-$N_c$, which provides a good approximation at $N_c=3$.\footnote{
Previously we argued the equivalence is restricted to a class of observables. As we will see, however, the equivalence holds for any observables. 
We thank F.~Karsch for a valuable critical comment, which made us revisit the issue and led to more precise statement. 
} 
The equivalence is a version of the large-$N_c$ orbifold equivalence \cite{Kachru:1998ys,Bershadsky:1998cb}, which was discovered 
through the study of the string theory. (For other interesting applications of the orbifold equivalence see e.g. \cite{Schmaltz:1998bg}.) 
In this paper, we briefly summarize the equivalence, show the lattice data which proves the equivalence can be seen already at $N_c=3$, 
and point out the same equivalence holds for various effective models in the mean field approximation.  

\section{The (partial) equivalence between QCD$_B$ and QCD$_I$ in the large-$N_c$ QCD}
\subsection{The equivalence between QCD$_B$ and QCD$_I$
}
Let us start with introducing the orbifold equivalence.  
First we choose the discrete symmetry $P$ 
(subgroup of gauge, flavor, or spacetime symmetry)
of the {\it parent} theory, which is the SO$(2N_c)$ or Sp$(2N_c)$ theory with the baryon chemical potential (SO$_B$ or Sp$_B$) in the present case. 
We then throw away all the degrees of freedom not invariant under $P$. This procedure is called the 
{\it orbifold projection}. After the projection, 
we obtain a new theory called the {\it daughter}.  
We consider two different projections, which give QCD with the baryon and isospin chemical potentials (QCD$_B$ and QCD$_I$) as daughters. 
The orbifold equivalence states that, in the large-$N_c$ limit,
correlation functions of operators ${\cal O}^{(p)}(A_{\mu},\psi)$
invariant under $P$ in the parent (called {\it neutral} operators)
agree with those of the operators 
${\cal O}^{(d)}(A_{\mu}^{\rm proj},\psi^{\rm proj})$
that consist of projected fields in the daughter: 
\begin{eqnarray}
\langle{\cal O}_1^{(p)}{\cal O}_2^{(p)}\cdots\rangle_{p}
=
\langle{\cal O}_1^{(d)}{\cal O}_2^{(d)}\cdots\rangle_{d}.
\end{eqnarray}
Here we take the coupling constants as 
$g_{\rm SU}^2=g_{\rm SO}^2=g_{\rm Sp}^2$,  
where the 't Hooft coupling $g_{\rm SU}^2N_c$ is kept finite. 
The field theoretic proof was given in \cite{Bershadsky:1998cb} for a class of theories, 
which can be generalized to various cases. 
For QCD$_B$, QCD$_I$, SO$_B$ and Sp$_B$, a couple of evidences of nonperturbative equivalence were also provided by 
the weak-coupling analysis in QCD and QCD-like theories 
at high density limit \cite{Hanada:2011ju}, 
low-energy effective theories \cite{Cherman:2011mh}, 
chiral random matrix models \cite{Hanada:2011ju} and holographic models \cite{Hanada:2012nj}. 

In order to build a projection from SO$_B$ to QCD$_B$, 
we use the ${\mathbb Z}_4$ discrete symmetries of SO$_B$  
generated by $J_c = -i\sigma_{2} \otimes 1_{N_{c}}$ 
($1_{N}$ is an $N \times N$ identity matrix) 
and $\omega = e^{i \pi/2} \in U(1)_B$.   
We require the gauge field $A^{\rm SO}_{\mu,ab}$ 
and the fermion $\psi^{\rm SO}_{\alpha,a}$ to be invariant under the following ${\mathbb Z}_2$ transformation 
embedded in the gauge and U$(1)_B$ transformation 
\cite{Cherman:2010jj},
\begin{eqnarray}
\label{eq:gauge}
A^{\rm SO}_{\mu,ab} &=& (J_c)_{aa'} A^{\rm SO}_{\mu,a'b'} (J_c^{-1})_{b'b}, \\
\label{eq:fermion_baryon}
\psi^{\rm SO}_{\alpha,a} &=& \omega (J_c)_{aa'} \psi^{\rm SO}_{\alpha,a'}. 
\end{eqnarray} 
From these projection conditions, it turns out that the daughter is QCD$_B$. 
The projection symmetry breaks down in the BEC/BCS crossover region (diquark condensation region) of SO$_B$, 
because the U$(1)_B$ symmetry is broken to ${\mathbb Z}_2$ there. 

One can also construct the projection from SO$_B$  
to QCD$_I$ for even $N_f$ by choosing another ${\mathbb Z}_2$ symmetry \cite{Cherman:2010jj,Hanada:2011ju},
\begin{eqnarray}
A^{\rm SO}_{\mu,ab} &=& (J_c)_{aa'} A^{\rm SO}_{\mu,a'b'} (J_c^{-1})_{b'b}, \\
\label{eq:fermion_isospin}
\psi_{\alpha,af}^{\rm SO} &=& (J_c)_{aa'} \psi_{\alpha,a'f'}^{\rm SO} (J_{i}^{-1})_{f'f},
\end{eqnarray}
where $J_{i} = - i\sigma_2 \otimes 1_{N_f/2}$ generates ${\mathbb Z}_4$ subgroup
of ${\rm SU}(2)$ isospin symmetry and the projection condition 
for the gauge field is the same as (\ref{eq:gauge}). 
In this case, the isospin symmetry used for the projection is unbroken everywhere, and so the 
orbifold equivalence holds including the BEC/BCS region of the phase diagram.
Therefore, through the equivalence with SO$_B$, 
we obtain the equivalence between QCD$_B$ and QCD$_I$ 
outside the BEC/BCS region of the latter; 
the phase quenching is exact for neutral sectors in this region.

Let us consider the $1/N_c$ corrections to QCD$_B$ and QCD$_I$. 
In the 't Hooft large-$N_c$ limit, 
expectation values of gluonic operators trivially agree because 
the fermions are not dynamical.
Now consider finite-$N_c$, say $N_c=3$ and $N_f=2$. 
Then the largest correction to the 't Hooft limit comes from 
one-fermion-loop planar diagrams, which, as we have seen, 
do not distinguish $\mu_B$ and $\mu_I$. 
Therefore the difference of expectation values of gluonic operators 
is at most $(N_f/N_c)^2$ (two-fermion-loop planar diagrams). 
In particular, the deconfinement temperatures, which are determined by 
the Polyakov loop, agree up to corrections of this order. A similar observation was made 
in \cite{Toublan:2005rq} by a perturbative argument. 
\subsection{More equivalence between QCD$_B$ and phase-quenched QCD}\label{QCDBvsPhaseQuench}
It is often said that QCD$_I$ and the phase quenched QCD are the same. 
However, although this statement is correct for the partition function, 
there is a difference when one considers the physical observables.   
In QCD$_I$, the propagators of up and down quarks are $D^{-1}(+\mu)$ and  $D^{-1}(-\mu)$. 
On the other hand, in the phase-quenched QCD, it is natural to take both $D^{-1}(+\mu)$.  
(In the terms of the lattice QCD simulation, the configuration are generated by using QCD$_I$, while  
the same operators as QCD$_B$ are used for the measurement.) 
Therefore the expectation values of the chiral condensate, the baryon density, and the isospin density are calculated as follows:   
\begin{eqnarray}
\begin{array}{|c|c|c|c|}
\hline
 & {\rm QCD}_B & {\rm QCD}_I & {\rm phase\mathchar`-quenched\ QCD}  \\
\hline
{\rm chiral\ condensate} &2 \langle {\rm Tr}D^{-1}(\mu)\rangle_B 
&  \langle {\rm Tr}D^{-1}(\mu)+ {\rm Tr}D^{-1}(-\mu)\rangle_I
& 2 \langle {\rm Tr}D^{-1}(\mu)\rangle_I\\ 
\hline
{\rm baryon\ density} &2 \langle {\rm Tr}\gamma^0D^{-1}(\mu)\rangle_B 
&  \langle {\rm Tr}\gamma^0D^{-1}(\mu)+ {\rm Tr}\gamma^0D^{-1}(-\mu)\rangle_I
&  2\langle {\rm Tr}\gamma^0D^{-1}(\mu)\rangle_I\\ 
\hline
{\rm isospin\ density} & 0
&  \langle {\rm Tr}\gamma^0D^{-1}(\mu)- {\rm Tr}\gamma^0D^{-1}(-\mu)\rangle_I
&  0\\
\hline
\end{array} 
\nonumber
\end{eqnarray}
Here $\langle\ \rangle_B$ and $\langle\ \rangle_I$ are the expectation values with QCD$_B$ and QCD$_I$ ensembles, respectively.  
We can easily see the chiral condensate in QCD$_I$ and the phase-quenched QCD take the same because of the charge-conjugation invariance of the QCD$_I$ ensemble. 
Therefore, the orbifold equivalence (the chiral condensate in QCD$_B$ = the chiral condensate in QCD$_I$) tells us 
it is not affected by the phase quenching. 
For the baryon density, let us remind $\langle {\rm Tr}\gamma^0D^{-1}(-\mu)\rangle_I=- \langle {\rm Tr}\gamma^0D^{-1}(+\mu)\rangle_I$, again because of the charge-conjugation invariance of 
the QCD$_I$ ensemble. Therefore, the isospin density in QCD$_I$ and the baryon density in the phase-quenched QCD take the same value. 
(Also the baryon density in QCD$_I$ becomes zero.) 
By combining it with the orbifold equivalence (the baryon density in QCD$_B$ = the isospin density in QCD$_I$), 
we conclude that the phase quenching does not affect the expectation value of the baryon density. 
The same argument holds for other observables too, and the orbifold equivalence leads to the exactness of the phase quenching for any observable to $O(N_f/N_c)$.  

Let us provide a heuristic argument which can be generalized to any number of flavors and any value of the quark chemical potentials. 
Let us assume that the overlap problem is not severe, as the argument based on the orbifold equivalence shows,  
and consider why the determinant phase does not modify the expectation values. 
In the large-$N_c$ limit, the quantum fluctuation is suppressed. In the phase quenched simulation, with a standard 't Hooft counting, 
properly normalized operators like $\bar{\psi}\psi/N_c$ fluctuates only $O(N_f/N_c)$. In other words, the histograms of the properly normalized 
quantities have a very sharp peak of the width $O(N_f/N_c)$. 
On the other hand, the phase factor is of order one\footnote{
Because the expectation value of the phase factor $\langle e^{i{\rm Im}S}\rangle$ is real, 
the one-point function $\langle {\rm Im}S\rangle$ is zero, 
and hence the leading contribution to the average phase comes from the connected two-point function of the imaginary part of the action, 
$\langle ({\rm Im} S)^2\rangle_{conn.}$, which is of order one. 
We thank Y.~Hidaka for a clear explanation on this point. 
}, and it cannot change drastically around the peak. 
Therefore the correction is at most of order $N_f/N_c$. 

\section{Evidence from lattice simulations at $N_c=3$}
We have seen that the phase quenching is exact to $O(N_f/N_c)$. 
But the standard 't Hooft counting does not tell us the expansion coefficients. 
That motivates us to look at lattice data of $N_c=3$ QCD. 
In the following we summarize lattice studies which compared QCD$_B$ and the phase-quenched QCD. 

\begin{itemize}
\item
In \cite{Nakamura:2005wg}, Nakamura et al. studied two-flavor QCD$_B$ and phase-quenched QCD 
by using staggered fermions. 
The QCD$_B$ is obtained by the phase reweighting. 
The bare quark mass is $am=0.05$
on a $8^3 \times 4$ lattice. 
The chiral condensate and the Polyakov loop are computed for $a\mu=0.1$ and $0.2$.  
They found a perfect agreement between QCD$_B$ and the phase-quenched QCD within numerical errors. 

\item
In the right panel of Fig.~1 and the left panel of Fig.~4 of \cite{deForcrand:2007uz}, 
the free energy at various temperatures between $0.5 T_c$ and $1.1 T_c$ are plotted 
as functions of $Q$. 
By putting these plots on top of each other, 
one can see a very nice agreement near the critical temperature 
and $Q \lesssim 100$. It clearly shows the validity of the phase quenching. 
It should also be remarked that the corrections are still tiny for $N_f=8$, 
a larger number of flavors than $N_f=2+1$ in the real world. 

\item
Fodor et al. \cite{Fodor:2007vv} combined the phase reweighting and the density of states methods. 
In Fig.~4 of \cite{Fodor:2007vv} they show the critical couplings at $a\mu=0.3$ 
both in the phase quenched and phase reweighted cases, which take close values.  

\item
The large-$N_c$ equivalence holds for the imaginary baryon and isospin chemical potentials, 
$(\mu_u,\mu_d)=(i\mu_{\rm img},i\mu_{\rm img})$ and 
$(\mu_u,\mu_d)=(i\mu_{\rm img},-i\mu_{\rm img})$, without any modification. 
As a result, the chiral condensates $\langle\bar{\psi}\psi\rangle_B$ and $\langle\bar{\psi}\psi\rangle_I$ 
take the same value at finite imaginary potentials as long as the projection symmetries are unbroken.

In \cite{Cea:2012ev}, the chiral critical temperatures $T_c(\mu)$ in two-flavor QCD 
were exploited by the extrapolations from the imaginary chemical potential, by using a fitting ansatz 
\begin{eqnarray}
\frac{T_c(\mu)}{T_c(0)}=1 + a_1 \left( \frac{\mu}{\pi T} \right)^2. 
\end{eqnarray}
They found \cite{Cea:2012ev}
$a_1 = -0.470(13)$ for $\mu_I$ and  
$a_1 = -0.522(10)$ for  $\mu_B$, 
which provide a nice quantitative agreement already at $N_c=3$. 

\item
Let us consider  the Taylor expansion method, in which   
the expectation value of an observable is expanded in powers of $\mu/T$, 
\begin{eqnarray}
\langle{\cal O}\rangle_{B,I}=\sum_{n=0}^\infty c_n^{B,I}\left(\frac{\mu}{T}\right)^n. 
\end{eqnarray} 
Taylor coefficients $c_n^B$ and $c_n^I$, which are functions of the temperature $T$,  
can be determined by the simulation at $\mu=0$. The large-$N_c$ equivalence tells 
that the coefficients agree in the large-$N_c$ limit: $\lim_{N_c\to\infty}c_n^{B} = \lim_{N_c\to\infty}c_n^{I}$.  

In \cite{Allton:2005gk}, the coefficient $c_2^B$ and $c_2^I$ for the chiral condensate 
and the pressure of the quark-gluon gass have been calculated\footnote{ For odd $n$, $c_{n}^B$ 
and $c_{n}^I$ vanish, and the first nontrivial $\mu$-dependences appear in 
$c_2^B$ and $c_2^I$. Although $c_{n}^B$ ($n \geq 4$) have been calculated, 
$c_{n}^I$ ($n\ge 4$) have not been calculated in \cite{Allton:2005gk}. 
(Note that, for $n\ge 4$, they use the same symbol $c_n^I$ for another quantity.)} 
in two-flavor QCD. 
Although the difference between $c_2^B$ and $c_2^I$ are not very small
for $T<T_c$ in the chiral symmetry broken (and confinement) phase, 
they agree exceptionally well for $T\gtrsim T_c$. 
The origin of the difference for $T<T_c$
may come from the contributions of thermally excited mesons which large-$N_c$ is suppressed at large-$N_c$. 
On the other hand, for $T>T_c$, fundamental degrees of freedom are
deconfined quarks and gluons rather than baryons or mesons, 
where the difference between QCD$_B$ and QCD$_I$ becomes much smaller 
and the large-$N_c$ equivalence is very well satisfied even at $N_c=3$.
  
\end{itemize}

\section{The equivalence in the effective models}
As an example of the effective models, we consider the Nambu--Jona-Lasinio (NJL) model. 
In order to simplify the discussion, we consider the chiral limit. 
The starting point is the Lagrangian with the U$(N_c)$ color current 
interaction with $N_f$ flavors,
\begin{eqnarray}
\label{eq:NJL_current}
{\cal L}_{\rm NJL}
=\bar{\psi}_f\left(
\gamma^\mu\partial_\mu + \mu_f\gamma^4
\right)\psi_f
+
\frac{G}{N_c} J^{({\rm U})}_{\mu A} J^{({\rm U})}_{\mu A}, 
\end{eqnarray}
where $J_{\mu A}^{({\rm U})} = \bar{\psi}_f\gamma_\mu T_{{\rm U}}^A\psi_f$ and 
$T^A_{{\rm U}}$ are the U$(N_c)$ color generators and 
summation is taken over repeated indices. 
The coupling constant $G$ is taken to be of order $N_c^{0}$.
One rewrites it keeping only the interactions in the 
scalar and pseudoscalar channels after Fierz transformations:
\begin{eqnarray}
{\cal L}_{\rm NJL}
&=&
\bar{\psi} \left(\gamma^\mu \partial_\mu + \mu_f\gamma^4 \right) \psi + {\cal L}_{\rm int},
\nonumber\\
\label{eq:4-fermi}
{\cal L}_{\rm int}&=& 
\frac{G}{N_c} \left[
(\bar{\psi}_f\psi_{f'})(\bar{\psi}_{f'}\psi_{f}) + 
(\bar{\psi}_f i\gamma^5\psi_{f'})(\bar{\psi}_{f'} i\gamma^5\psi_{f})
\right]. 
\label{eq:NJL}
\end{eqnarray}
In the Lagrangians (\ref{eq:NJL_current}) and (\ref{eq:NJL}), 
the invariance under U$(N_c)$ gauge symmetry 
and U$(N_f)_L \times U(N_f)_R$ flavor symmetry are manifest.
Here we ignore the effect of instantons or the U$(1)_A$ anomaly
which explicitly breaks the U$(1)_A$ symmetry, because it is
subleading in $1/N_c$. 
(From the viewpoint of the orbifold equivalence, 
there is no reason for the exactness of the phase quenching
at the level of mean-field approximation (MFA) if we take into 
account the $1/N_c$-suppressed instanton effects.
However, even if we incorporate them, 
the phase quenching for the chiral condensate turns out to be exact within the NJL model \cite{Hanada:2012es}.)  
For SO$(2N_c)$ theory, we can construct the corresponding 
NJL model in the same manner, by using the SO$(2N_c)$ current. 
The proof of the large-$N_c$ orbifold equivalence applies to the NJL model, 
by starting with the NJL model for SO$_B$ and by using similar projection conditions as the previous section \cite{Hanada:2012es}. 

When one considers the large-$N_c$ limit, one should set up 
the correct $1/N_c$-counting scheme which reproduce the correct $1/N_c$-scaling in the large-$N_c$ QCD. 
The quark $\psi$ has $N_c$ colors so that a closed color loop gives a factor of $N_c$. 
The coupling constant of the four-fermi interaction should be taken as $O(N_c^{-1})$, 
and furthermore, 
the form of possible four-fermi interactions are restricted; in other words only the interactions 
which have origins in QCD are allowed. 
In this setup, the right $1/N_c$-counting follows  and we can use the same proof
of the orbifold equivalence as the large-$N_c$ QCD \cite{Hanada:2012es}.

Now let us see the relationship between the large-$N_c$ limit and the MFA. 
We first perform the Hubbard-Stratonovich transformation
by introducing auxiliary fields corresponding to the fermion bilinears,
$\sigma = \bar \psi \psi$ and $\pi_a = \bar \psi i \gamma_5 \tau_a \psi$, 
and then integrate out fermions to obtain the partition funcion
\begin{eqnarray}
Z \equiv e^{-W} = \int d \sigma d \pi e^{-I(\sigma, \pi)}.
\end{eqnarray}
Here $I(\sigma, \pi)$ is the bosinized effective action 
\begin{eqnarray}
I(\sigma, \pi) = N_c \left[-{\rm {\rm Tr}} \log D + \frac{1}{G}\int d^4 x (\sigma^2 + \pi_a^2)\right],
\end{eqnarray}
with $D=\gamma^\mu\partial_\mu+2\sigma_A+2\pi_A$. 
It describes mesons $\sigma$ and $\pi$.
Because of an overall factor $N_c$ in the action, the $1/N_c$-expansion is equivalent to the expansion 
with respect to meson loops. 
The leading order corresponds to the saddle-point approximation, 
or equivalently the conventional MFA where the auxiliary fields are 
replaced by the expectation value (i.e., the mean-field). 
In order to go beyond the MFA, we have to take into account
meson-loops order by order. 

Similar arguments hold also for various other theories, 
such as linear sigma model \cite{Hanada:2012es}, Polyakov-Nambu-Jona-Lasinio model \cite{Hanada:2012es}, Polyakov-quark-meson model \cite{Hanada:2012es}, 
chiral random matrix model \cite{Hanada:2011ju}, Sakai-Sugimoto model \cite{Hanada:2012es} and D3/D7 model \cite{Hanada:2012nj}.

\section{Conclusion and Outlook}
We have seen the phase quenching is exact to $O(N_f/N_c)$, outside the pion condensation of the phase quenched theory. 
In other words, the effect of the phase is $1/N_c$-suppressed, and hence the reweighting method works without being threatened 
by the overlap problem. Previous lattice studies confirm the effect of the phase is small already at $N_c=3$. 
We have also shown the exactness of the phase quenching in effective models, which had been realized by explicit calculations and used to 
justify the reweighting method, can be understood in a unified manner, from the point of view of the large-$N_c$ equivalence. 
Now the phase quench and phase reweighting methods have a theoretical justification, and hence it is important to study the QCD phase diagram by using them.

\end{document}